\documentclass[acus]{JAC2003}

\usepackage{graphicx}
\usepackage{booktabs}

\begin{document}
\title{COMPENSATION OF TRANSVERSE FIELD ASYMMETRY IN THE HIGH-BETA QUARTER-WAVE RESONATOR OF THE HIE-ISOLDE LINAC AT CERN%\thanks{The author (M.A. Fraser) acknowledges the receipt of funding from the ISOLDE Collaboration Committee and the Cockcroft Institute.}
}
\author{M.A. Fraser$^{\dag \ddag}$\thanks{matthew.alexander.fraser@cern.ch}, M. Pasini$^{\S \ddag}$, A. D'Elia$^{\dag \ddag}$, R.M. Jones$^{\dag}$\\
$^{\dag}$The University of Manchester, Oxford Road, Manchester, M13 9PL, UK.\\
$^{\dag}$The Cockcroft Institute, Daresbury, Warrington, Cheshire WA4 4AD, UK.\\
$^{\S}$Instituut voor Kern- en Stralingsfysica, K.U.Leuven, Celestijnenlaan 200D
B-3001 Leuven, BE.\\
$^{\ddag}$CERN, Geneva, Switzerland.\\}

\maketitle

\begin{abstract}
   The superconducting upgrade of the REX-ISOLDE radioactive ion beam (RIB) post-accelerator at CERN will utilise a compact lattice comprising quarter-wave resonators (QWRs) and solenoids, accelerating beams in the mass range 2.5 $<$ A/q $<$ 4.5 to over 10 MeV/u. The short and independently phased quarter-wave structures allow for the acceleration of RIBs over a variable velocity profile and provide an unrivalled longitudinal acceptance when coupled with solenoid focusing. The incorporation of the solenoids into the cryomodule shortens the linac, whilst maximising the acceptance, but the application of solenoid focusing in the presence of asymmetric QWR fields can have consequences for the beam quality. The rotation of an asymmetric beam produces an effective emittance growth in the laboratory reference system. We present modifications of the cavity geometry to optimise the symmetry of the transverse fields in the high-$\beta$ QWR. A racetrack shaped beam port is analysed and a modification made to the inner conductor with a geometry that will enable a niobium film to be effectively sputtered onto the cavity surface.
\end{abstract}

\section{INTRODUCTION}

\begin{figure*}[ht]
   \centering
   \includegraphics*[width=160mm]{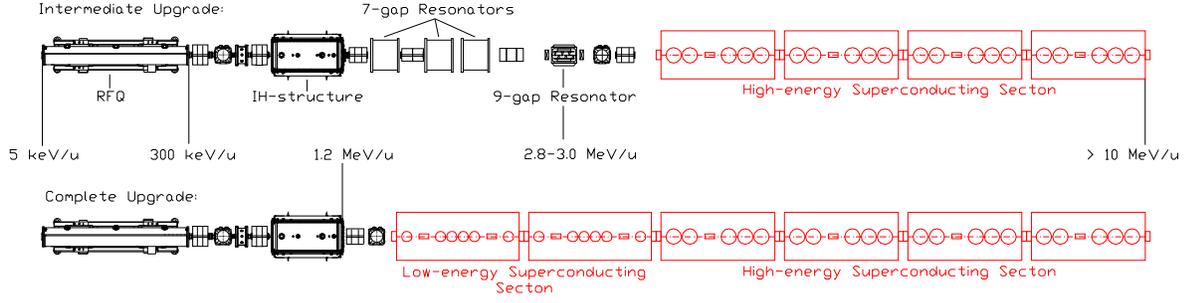}
   \caption{The layout of two stages of the HIE-ISOLDE linac upgrade: intermediate (top) and complete (bottom). The existing normal conducting infrastructure is shown in black and the new superconducting upgrade is shown in red.}
   \label{layout}
\end{figure*}

The energy upgrade of RIBs at CERN under the HIE-ISOLDE framework will be realised using a superconducting linear post-accelerator comprising quarter-wave resonators of two geometries, corresponding to low and high energy with reduced velocities of 6.3~\% and 11.3~\%, respectively \cite{tech_opts}. The beam will be focused using superconducting solenoids located within the cryomodules in order to maximise both the transverse and longitudinal acceptance of the machine and to allow for easy scaling of linac settings when accelerating radioactive beam. In total, twelve low-$\beta$ cavities  and four solenoids will be housed in the first two cryomodules providing an energy of at least 3.6~MeV/u at injection into the high energy section, which will contain twenty cavities and four solenoids housed in 4 cryomodules. Two stages of the planned upgrade of REX-ISOLDE are shown in  Figure~\ref{layout}. This paper focuses on the high-$\beta$ cavity and consequently the beam dynamics of the high energy section of the HIE-ISOLDE linac. The study presented analyses the effect of the asymmetric rf fields of the QWR on the beam and investigates two modifications of the nominal cavity to improve the symmetry of the fields: The first a simple modification to the beam port and the second a redesign of the drift tube on the internal conductor.

\begin{figure}[htb]
   \centering
   \includegraphics*[width=60mm]{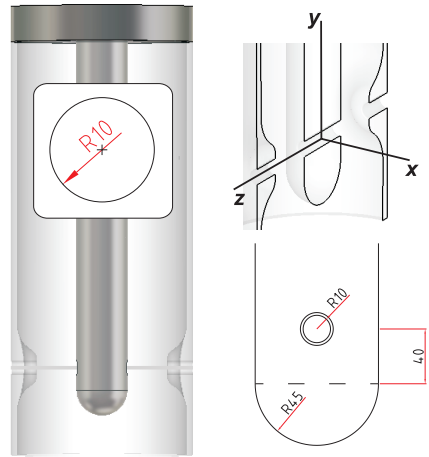}
   \caption{The nominal high-$\beta$ cavity (left), the co-ordinate system (top right) and the geometry of the internal conductor (bottom right). Dimensions in mm.}
   \label{hb}
\end{figure}

\section{THE HIGH-$\beta$ CAVITY}

\begin{figure}[htb]
   \centering
   \includegraphics*[width=70mm]{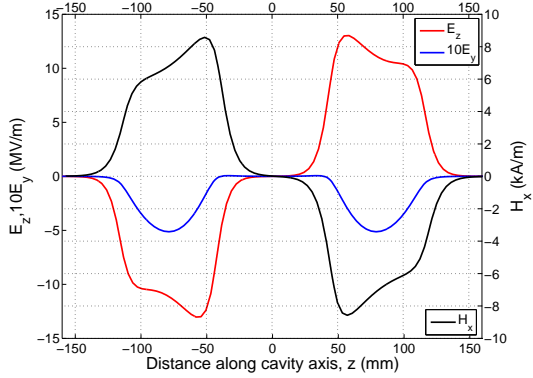}
   \caption{The longitudinal and transverse field components on the beam axis of the high-$\beta$ QWR. \textit{H$_{\textrm{x}}$} is shown with a 90$^{\circ}$ phase delay to \textit{E}. \textit{E$_{\textrm{y}}$} is enhanced by a factor of 10.}
   \label{fields}
\end{figure}

The high-$\beta$ cavity has a design gradient of 6~MV/m, providing 1.8~MV of accelerating potential over an active length of 30~cm, at a resonant frequency of 101.28 MHz. The cavity will be made from a copper substrate onto which a niobium film will be sputtered, details of which can be found in \cite{cav}. The cavity geometry is shown in Figure~\ref{hb} as drawn in the \texttt{MWS} code used to numerically simulate the rf fields of the cavity \cite{MWS}.  The nominal cavity design has a circular aperture of 20~mm, instead of the racetrack variant, as presented, inset, in Figure~\ref{variants}. The cavity noses are formed by pressing the external cylindrical copper sheet. More information regarding the cavity design and manufacture can be found in \cite{hb_cavity}. The TEM modes characteristic of this type of cavity geometry introduce an appreciable component of magnetic field on axis, as shown in Figure~\ref{fields}.

\begin{figure}[htb]
   \centering
   \includegraphics*[width=70mm]{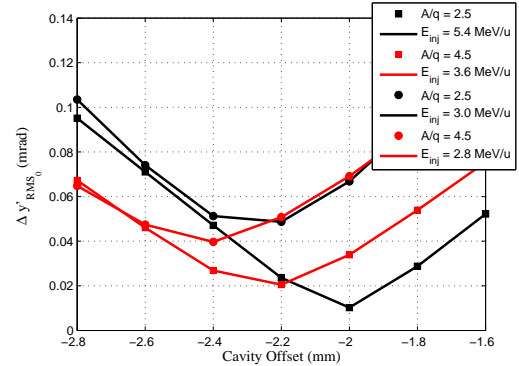}
   \caption{The optimisation of cavity offset for the nominal cavity.}
   \label{offset}
\end{figure}

The beam steering effect is phase and energy dependent and a source of coupling between the transverse and longitudinal dynamics, which can cause significant transverse emittance dilution if not compensated. The method of compensation employed in the high-$\beta$ cavity requires an offset of the cavity relative to the beam, such that the beam centroid samples the electric (de)focusing field created by the beam port, providing good compensation over a wide velocity range, as shown in Figure~\ref{graphs}. The compensation scheme is detailed in \cite{beam_steer}. The velocity dependency of the steering force and the variable velocity profile along the linac for different mass-to-charge states makes the optimum cavity offset dependent on A/q. The offset of the cavities was determined by minimising the sum of the squares of the steering kicks in each cavity along the linac, across the mass-to-charge state acceptance, i.e.~from A/q = 2.5 to 4.5, as shown in Figure~\ref{offset}. A compromise was made to keep the deflection low over all RIBs in the machine's acceptance. In fact, the increased sensitivity of low mass-to-charge RIBs to the steering force is compensated somewhat by the higher velocity developed in the linac. The range of the velocity profiles in the high energy section is shown in Table~\ref{energy}, along with the optimised offset for the nominal cavity and its racetrack variant, at injection energies provided by the existing REX-ISOLDE linac and by the low energy superconducting section in the complete upgrade. The offset is dependent on the aperture geometry. In the nominal case with no misalignment, the centroid excursion can be kept to below 0.5~mm along the entire 10.5~m length of the high energy section, as shown in the bottom series of graphs in Figure~\ref{bd}.

 \begin{table}[hbt]
   \centering
  \footnotesize
   \caption{The range of the energy, $E$, velocity, $\beta$ and the corresponding optimum cavity offset, at injection after the 9-gap resonator (top series) and after the low energy superconducting section (bottom series).} 
   
   \begin{tabular}{cp{2cm}p{1.5cm}p{1.5cm}p{0mm}}
       \toprule
       \textbf{A/q} &\centering \textbf{\textit{E}$_{inj}$/$\mathbf\beta$ (MeV/u)/(\%)} &\centering \textbf{\textit{E}$_{ej}$/$\mathbf\beta$ (MeV/u)/(\%)} &  \centering  \textbf{Offset (NOM/RT) (mm)} &\\ 
       \midrule 
        4.5     &  \centering2.8 / 7.8 &  \centering  9.3 / 14.0    & \centering  2.4 / 3.0&     \\
           2.5      & \centering3.0 / 8.0 & \centering14.4 / 17.4  &  \centering 2.3 / 2.8&     \\
           \midrule
           4.5     &  \centering3.6 / 8.8 & \centering10.2 / 14.7    & \centering 2.2 / 2.8 &   \\
           2.5      &\centering 5.4 / 10.7 & \centering16.6 / 18.6  &    \centering  2.0 / 2.5 &    \\
         \bottomrule
   \end{tabular}
   \label{energy}
\end{table}

\section{MODIFIED CAVITY GEOMETRIES}

\begin{figure}[htb]
   \centering
   \includegraphics*[width=70mm]{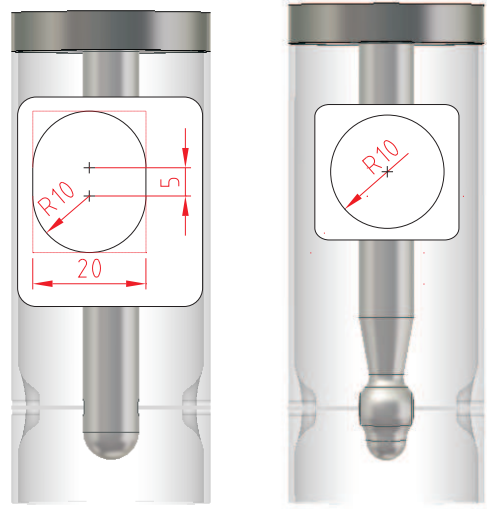}
   \caption{The RT cavity (left) and MOD DT cavity (right) (exported from \texttt{MWS}). Inset is the racetrack geometry with dimensions in mm.}
   \label{variants}
\end{figure}

Two cavity variants are presented in Figure~\ref{variants}: The nominal cavity with a racetrack aperture (RT) and with a modified drift tube (MOD DT). Their parameters are displayed for comparison with the nominal cavity (NOM), in Table ~\ref{para}. 

\begin{table}[hbt]
   \centering
   \small
   \caption{Comparison of Cavity Variants}
   \begin{tabular}{lccc}
       \toprule
       \textbf{Parameter} & \textbf{NOM} & \textbf{RT} & \textbf{MOD DT} \\ 
       \midrule
           TTF$_{max}$($\beta=\beta_{max}$)& 0.904            & 0.902 & 0.904            \\
           $\beta_{max}$ (\%)& 11.3            & 11.3 & 11.3             \\
           Mechanical Height (mm) & 785            & 785 & 807            \\
           Mechanical Length (mm) & 320            & 320 & 320             \\
           Active Length (mm) & 300           & 300 & 300            \\
           Drift Length (mm) & 90            & 90 & 90             \\
           Gap Length (mm) & 70            & 70 & 70            \\
           E$_{peak}$/E$_{acc}$     & 5.4            & 5.4 & 6.6            \\
           H$_{peak}$/E$_{acc}$ Oe/(MV/m)  & 96            & 96 & 93          \\
           R$_{shunt}$/Q$_{0}$ $(\Omega)$ & 548            & 548 & 585          \\
           $\Gamma = R_{surface}.Q_{0}$ $(\Omega)$ & 30.6            & 30.6 & 31.3          \\
           U/E$^2_{acc}$ J/(MV/m)$^2$  & 207            & 207 & 196         \\
           Compensation Offset (mm)  & 2.2            & 2.8 & 2.2          \\
         \bottomrule
   \end{tabular}
   \label{para}
\end{table}

\begin{figure}[htb]
   \centering
   \includegraphics*[width=70mm]{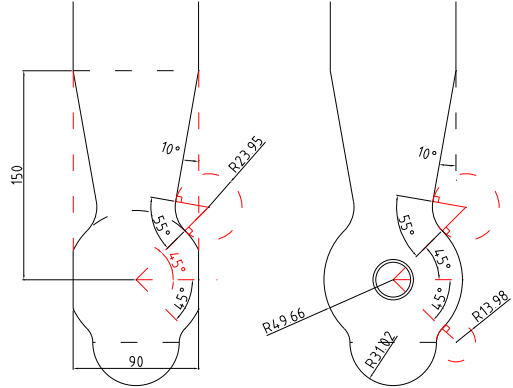}
   \caption{The design of the modification to the internal conductor. Dimensions in mm.}
   \label{DT}
\end{figure}

\begin{figure}[ht]
   \centering
   \includegraphics*[width=70mm]{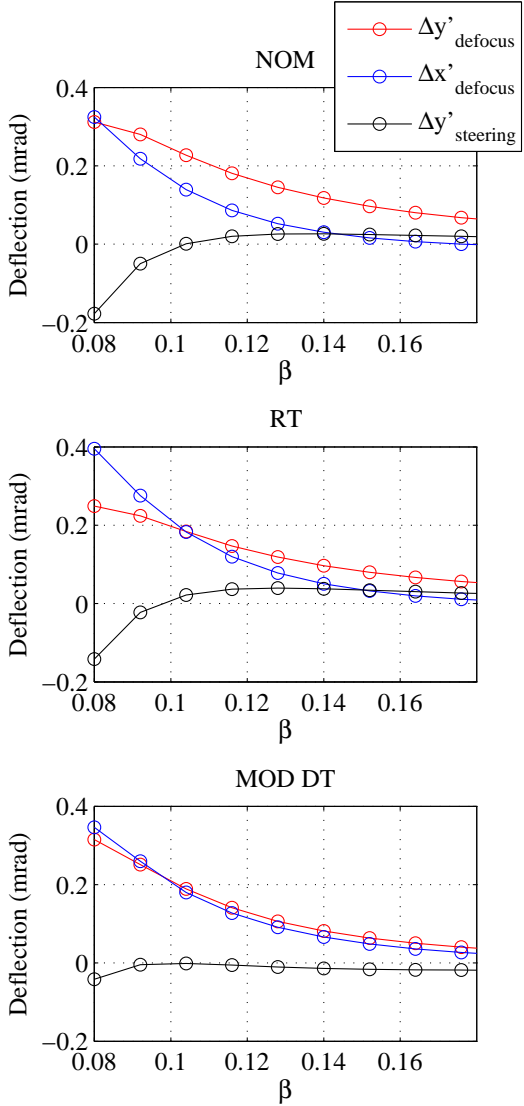}
   \caption{The rf defocusing force at 1 mm from the offset beam axis and beam steering force in the nominal high-$\beta$ cavity with a circular aperture (top), modified with a racetrack aperture (middle) and with a modified drift tube (bottom). A/q = 2.5 and $\phi_{s}$ = -20$^{\circ}$. }
   \label{graphs}
\end{figure}

\begin{figure}[ht]
   \centering
   \includegraphics*[width=80mm]{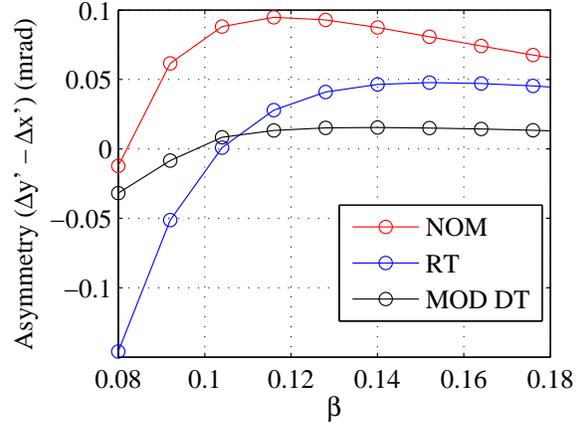}
   \caption{A comparison of the asymmetry in the rf defocusing force, arising from the quadrupolar component of the cavity fields on the beam axis. A/q = 2.5 and $\phi_{s}$ = -20$^{\circ}$.}
   \label{asym}
\end{figure}

The dimensions of the racetrack aperture are shown inset on Figure~\ref{variants}. The racetrack minimises the loss in aperture when the cavity is offset and keeps the shunt impedance high with respect to a circular beam port. The modification to the drift tube of the internal conductor reflects the shape of the noses and is achieved by tapering a flattened sphere onto the shaft. An extension to the bottom of the sphere is required to move the site of peak electric field away from the beam axis, such that an effective compensation of beam steering can be achieved. The modification to the internal conductor is shown in Figure~\ref{DT}. With the exception of the flattened faces, the modification has cylindrical symmetry about the resonator axis and would present an acceptable surface for sputtering. The beam dynamics performance of each cavity was analysed by numerically integrating the Lorentz equation of motion of a single particle in the electromagnetic fields exported from \texttt{MWS}. A mesh size of 1~mm was used and the field values interpolated linearly between the mesh at each integration step. A comparison of the rf defocusing force experienced by an ion of A/q = 2.5 at a synchronous phase of -20$^{\circ}$, 1~mm from the offset axis in the vertical and horizontal directions, is shown as a function of reduced velocity in Figure~\ref{graphs} for the three different cavities investigated. Also shown is the beam steering force. The asymmetry of the rf fields in each cavity variant, defined as the difference between the defocusing force in the vertical and horizontal directions at 1~mm from the offset beam axis, is summarised in Figure~\ref{asym}. One can see that over most of the velocity profile of interest for A/q = 2.5, the racetrack aperture reduces the asymmetry compared to the circular aperture. The asymmetry of the fields in the QWR is intrinsic to the cylindrical geometry and can be attributed to the way the cavity walls bend away from the beam port in the vertical and horizontal directions. The racetrack shape keeps the difference in the horizontal and vertical rf defocusing forces to less than 0.05~mrad for ions with A/q = 2.5, in the complete upgrade. 

The improvement with the racetrack shape arises by orientating the major axis vertically, which reduces the amount of metal seen above and below the aperture at the beam axis. Although the racetrack shape is beneficial on a cylindrical geometry, a circular aperture is required to optimise the field symmetry on the modified internal conductor. The fields are kept symmetric to better than 0.03~mrad across the entire velocity range with the modification made to the internal conductor. When the upgrade is completed the injection energy will be lifted to at least 3.6~MeV/u, making the intrinsic asymmetry of the racetrack variant preferable over the circular aperture. The modification to the internal conductor reduces the capacitive loading on the QWR and, in order to maintain the resonant frequency of 101.28 MHz, the cavity must be lengthened. The modification required to compensate for the field asymmetry causes the peak electric field to rise as a result of reducing the radius of curvature of the high field region on the end of the internal conductor.

\section{BEAM DYNAMICS SIMULATIONS}

\begin{figure*}[ht]
   \centering
   \includegraphics*[width=160mm]{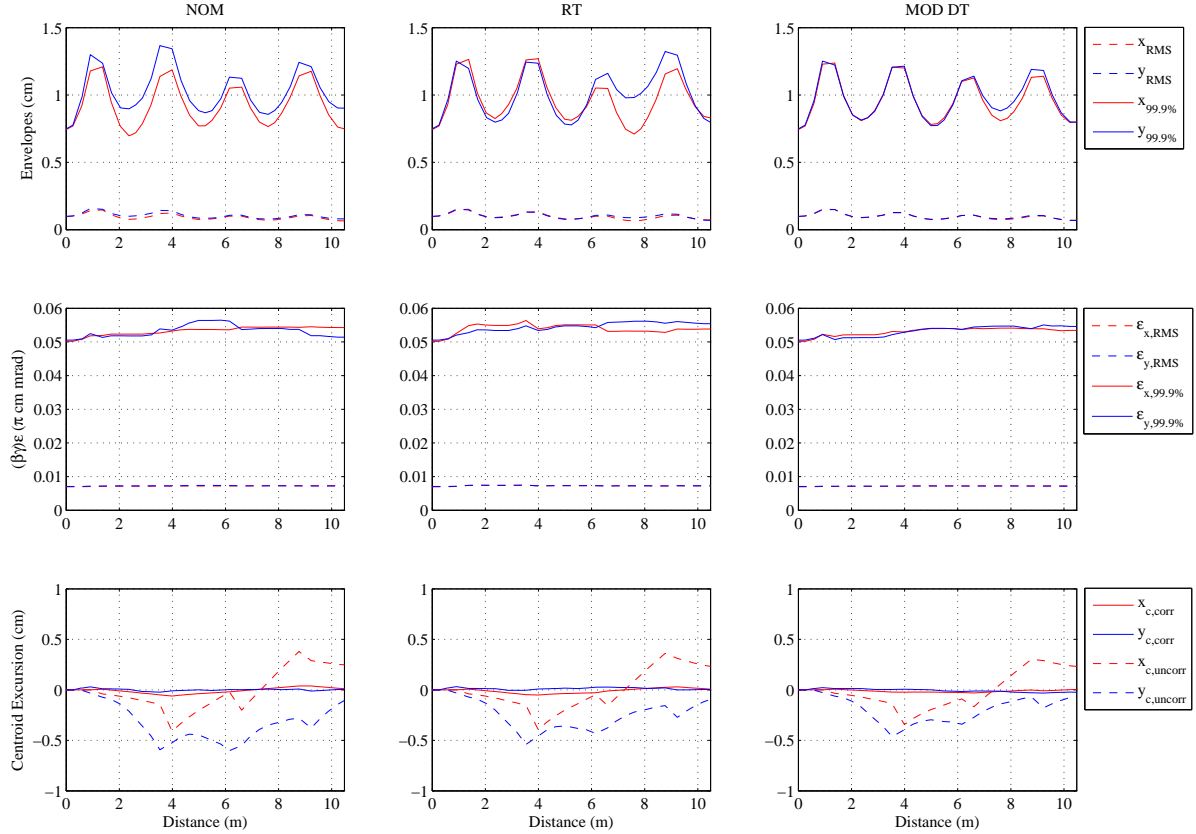}
   \caption{The salient characteristics of the beam for the three cavity geometries: circular aperture (left series), racetrack aperture (middle series) and the modified drift tube (right series).}
   \label{bd}
\end{figure*}

Multi-particle beam dynamics simulations of the linac containing the three different cavity variants were carried out using the \texttt{TRACK} code \cite{TRACK}. The linac was first tuned using the \texttt{LANA} code, in which the cavity and solenoid fields are represented by hard-edge models \cite{LANA}. The superconducting linac is operated with a focusing channel delivering a phase advance between solenoids of 90$^{\circ}$, chosen to ensure a resonant growth in emittance is suppressed, as reported in \cite{resonance}. The results presented in Figure~\ref{bd} are of the dynamics of the lightest beams with A/q of 2.5 at an injection energy of 3.0 MeV/u. The simulations are of the nominal linac without error. A comparison of the emittance evolution of the the three cavities indicates that the main source of transverse emittance growth comes from the phase dependency of the transverse rf defocusing and beam steering forces. From the trajectory of the beam centroid shown in the bottom series of graphs in Figure~\ref{bd} it is clear that the racetrack compensates equally as well for beam steering as the circular aperture. The asymmetric fields split the beam envelopes as a result of different phase advances in the horizontal and vertical planes, making the linac less tolerant to mismatch resulting from errors in the injected beam parameters and solenoid field values. The modification to the internal conductor removes the asymmetry and maintains beam envelopes of the same dimension. With these linac settings, along the first half of the linac the racetrack aperture keeps the envelope of the beam symmetric. As an asymmetric beam is rotated in the solenoid, the horizontal and vertical emittances are mixed, which results in an effective increase (or decrease) of the transverse emittance in the horizontal and vertical planes. The effect of the quadrupole component of the fields can be observed clearly by the structure, at the level of a few percent, in the evolution of the emittance at the solenoid positions along the linac. The modification to the internal conductor ensures that the splitting of the $x$ and $y$ emittance is minimised and that the increase in the emittance is smooth, however, the improvement to the beam quality is only of the order of a few percent.

\section{CONCLUSIONS}

We have investigated the field asymmetry intrinsic to the QWR by studying three different cavity geometries in the high energy section of the HIE-ISOLDE linac. A racetrack shaped aperture is shown to reduce the field asymmetry, with respect to a circular aperture, above an energy of 3.7~MeV/u ($\beta$ = 0.09) on the cylindrical geometry of the QWR. This energy coincides very closely with the lowest injection energy foreseen in the complete upgrade possessing the low energy superconducting section, as was shown in Table~\ref{energy}. A major modification to the internal conductor was shown to make the fields highly symmetric on the beam axis, however, the small improvement to the beam quality is achieved at the expense of a 20~\% rise in the peak electric field.

\section{ACKNOWLEDGMENTS}

The author acknowledges the receipt of funding from the ISOLDE Collaboration Committee and the Cockcroft Institute.

\end{document}